\documentclass[prl,twocolumn,showpacs,amsmath,amssymb,superscriptaddress]{revtex4}

\usepackage{amsmath,amssymb}
\usepackage{graphicx}
\usepackage{epsfig}     
\usepackage{dcolumn}
\usepackage{bm}

\begin{document}

\title{Optical Lattices for Atom Based Quantum Microscopy}
\author{Andreas Klinger}
\affiliation{James Franck Institute and Physics Department, The University of Chicago,
Chicago, Illinois 60637, USA}
\author{Skyler Degenkolb}
\affiliation{James Franck Institute and Physics Department, The University of Chicago,
Chicago, Illinois 60637, USA}
\author{Nathan Gemelke}
\affiliation{James Franck Institute and Physics Department, The University of Chicago,
Chicago, Illinois 60637, USA}
\author{Kathy-Anne Brickman Soderberg}
\affiliation{James Franck Institute and Physics Department, The University of Chicago,
Chicago, Illinois 60637, USA}
\author{Cheng Chin}
\affiliation{James Franck Institute and Physics Department, The University of Chicago,
Chicago, Illinois 60637, USA}


\date{\today}
\begin{abstract}
We describe new techniques in the construction of optical lattices to realize a coherent atom-based microscope, comprised of two atomic species used as target and probe atoms, each in an independently controlled optical lattice.
Precise and dynamic translation of the lattices allows atoms to be brought into spatial overlap to induce atomic interactions.
For this purpose, we have fabricated two highly stable, hexagonal optical lattices, with widely separted wavelengths but identical lattice constants using diffractive optics.
The relative translational stability of 12~nm permits controlled interactions and even entanglement operations with high fidelity.
Translation of the lattices is realized through a monolithic electro-optic modulator array, capable of moving the lattice smoothly over one lattice site in 11~$\mu$s, or rapidly on the order of 100~ns.
\end{abstract}
\maketitle
\section{1. Introduction}

Optical lattices have provided a basis for numerous recent
experiments with ultracold atoms, ranging from basic physics of
single atoms in
lattice potentials \cite{bloch_osc, wsl}, to interacting quantum gases \cite{bloch_mottins}, and promises extension into
quantum magnetism and exotic quantum phases \cite{OL_review}.
Particularly stimulating applications arise from the field of
quantum information, where ultracold atoms in optical lattices are
proposed to be good vehicles for the storage and processing of
information \cite{Jaksch1999, Brennen1999, jaksch2000, jessen2004, daley2008, Bloch2008, KA2009}. Carefully
engineered lattice potentials, even with dynamic modes of control,
have become more common in modern quantum gas experiments
\cite{Trotzky2007, Anderlini2007}. General methods for
introducing additional degrees of freedom to periodic optical
potentials to, for example, engineer the structure of a unit cell
\cite{Lundblad2008}, or trap and manipulate multiple atomic species
\cite{KA2009} are necessary to make further progress in these
fields. Furthermore, the ability to probe and manipulate atomic
information \cite{you2000, calarco04, shimizu04, STMcoldatoms,Wurtz2009, Steinhauer09, lundblad09} with a spatial resolution well below the lattice spacing
is of primary importance in these types of experiments \cite{nelson2007, Greiner09}.

\begin{figure}[h]
\begin{center}
\includegraphics[width=0.8\columnwidth,keepaspectratio]{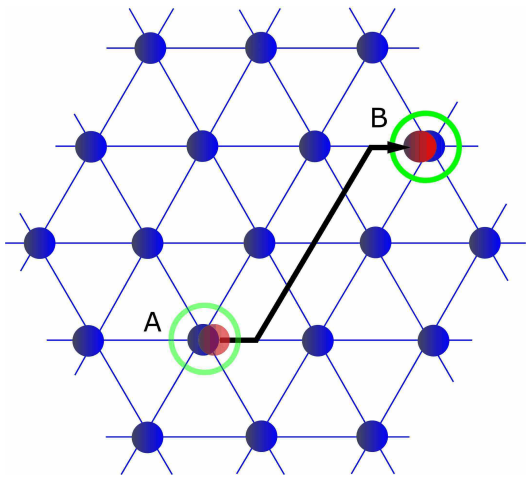}
\caption{Controlled overlap of two atomic species in two optical lattices with tunable translation. The target atoms
(blue circles) are confined in one lattice (blue lines), while
the probe atom (red circles) is confined by another lattice (not shown). The overlap of the probe and target atoms is controlled by shifting the optical phases of the lattice beams.  The figure shows an interaction between a target and probe atom at point A.  After the operation is finished, the probe atom can either have its state read by measurement, or alternatively transported to another target atom for a second quantum manipulation, as shown at point B.}
\label{generalscheme}
\end{center}
\end{figure}

We describe the construction of an apparatus to achieve these aims using a new
type of cold atom based microscopy, which we refer to as atomic
quantum microscopy.  This technique is a marriage of microscopy and quantum control, allowing the use of
one quantum object to probe and manipulate
a larger, more complex quantum system. Here, we employ two species of ultracold atoms,
each confined near the minima of an
optical lattice potential.  The lattice wavelengths and intensities are chosen to ensure that each atomic species can be fully controlled by its respective lattice \cite{KA2009}.
Through the combined use of controlled atomic collisions and spatial overlap of their wavefunctions, information
concerning internal states and external degrees of freedom can be
transferred from one species to the other, as illustrated in Fig. \ref{generalscheme}.  This apparatus makes possible new types of measurements, such as probing the full quantum structure
of complex many-body systems of interacting gases of atoms.  Since the probe can be entangled with the sample,
information can be read out in a way that does not interfere with the underlying more
complicated quantum system, permitting idealized projective quantum
measurements on a many-body system. In addition, entanglement between spatially separated portions of the sample can either be induced or interrogated.
Application of this apparatus for scalable quantum information processing was described previously \cite{KA2009}, where a lattice potential at one
wavelength confines qubit-atoms, and the probe atoms play the role
of auxillary messengers mediating entanglement among
(distant) qubits. Controlled interactions between species were
proposed to entangle the internal states of atoms brought into close
spatial contact.

We utilize novel techniques in the construction of optical lattices to produce
commensurate lattice spacings with widely separated laser wavelengths,
while providing precise differential control of the lattice site alignment
through manipulation of optical phases. A relative translational stability of 12~nm is demonstrated. Based on previous quantitative analysis of a practical lattice
configuration \cite{KA2009}, this yields an overlap fidelity of $>$ 99\%.

In this article, we describe novel optical and opto-electronic
components, including diffractive optical elements,
aberration-balanced imaging, a monolithic array of electro-optic
phase modulators, and the necessary driving electronics.  These methods
are applicable to both ultracold atom experiments exploring many-body physics and
quantum information processing, where optical
lattices with complex and/or dynamically manipulable unit cells are
desired. The techniques may also be exploited in the manipulation of microscale objects of biological or other interest \cite{Grier2003}.

\begin{figure}[h]
\begin{center}
\includegraphics[width=1.0\columnwidth,keepaspectratio]{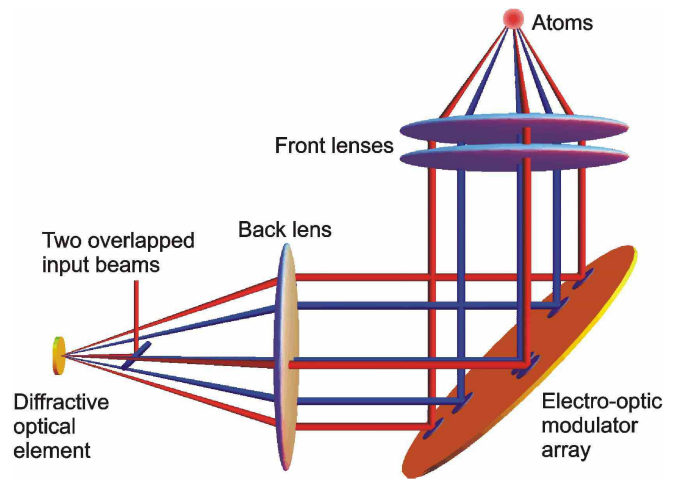}
\caption{Method for generation of a bichromatic optical lattice. Two
overlapped beams of wavelength $\lambda_1$ and $\lambda_2$ are incident
on the diffractive optical element. The three primary diffracted orders at each wavelength are collimated by the the back lens, phase-modulated with the electro-optic modulator array,
and focused onto the atoms by a pair of lenses.  The angles of recombination are naturally matched to produce commensurate lattices.}
\label{setup}
\end{center}
\end{figure}

\section{2. Commensurate, Multi-Chromatic Optical Lattices}
\label{lattices}

Optical lattices are formed by overlapping two or more laser beams
with wavenumbers $k=2\pi/\lambda$ on the atomic
sample. When two beams intersect with an angle $\Delta\theta$,
they create a sinusoidal interference pattern with spatial
frequencies $k'=k/(2\sin \Delta\theta)$, and phases determined by
the relative optical phases of the beams.  Here our goal is to construct bi- or multi-chromatic
optical lattices with extreme relative phase stability and
commensurate lattice constants for quantum microscopy.  We
achieve these goals by forming two lattices holographically, employing only
common-mode optics.

The optics for a hexagonal lattice are schematically shown
in Fig.~\ref{setup}; this is the simplest two-dimensional geometry
which is topologically stable against drifts in relative optical
phases \cite{jessen96}. Two co-propagating beams with separated wavelengths
$\lambda_1$ and $\lambda_2$ are incident on the two-dimensional
reflective diffraction grating shown in Fig.~\ref{afm_image},
splitting each primarily into three first-order diffracted beams.
After collimation by a lens, the diffracted beams pass
through an electro-optic modulator (EOM) crystal (see
Sec. 3), which imparts controlled phase differences
between beams.  They are then recombined and focused onto the
atoms by a compound lens. The
wavelength dependence of the diffraction angle at the diffractive optic naturally matches the angles of recombination, such that the
resultant lattice geometries are identical in the absence of lens aberration (discussed in Sec. 2B).  A separate one-dimensional lattice potential (not shown in the figure), applied along the optical axis and common to both atomic species, is used to complete the three-dimensional lattice potential.

\subsection{A. Diffractive Optics}
\label{holographic_optics}

\begin{figure}[h]
\begin{center}
\includegraphics[width=0.75\columnwidth,keepaspectratio]{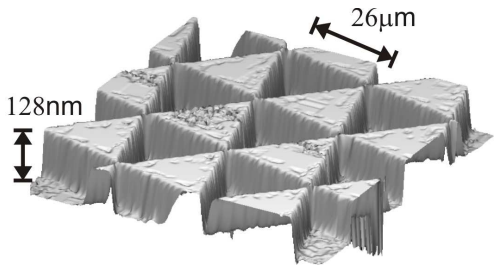}
\caption{Atomic force microscope image of a diffractive optical
element used to form the bichromatic lattice.  The substrate is fused-silica, patterned by photolithography, and overcoated with gold.
A reflected laser beam receives a differential phase shift due to propagation delay into the recessed areas, forming a triangular diffraction pattern in the far-field.}
\label{afm_image}
\end{center}
\end{figure}

The central element in producing the bichromatic lattice above is
the two-dimensional diffractive optical element (DOE) shown in
Fig.~\ref{afm_image}, which imprints a patterned phase shift across
the profile of the incident beam to create a
two-dimensional diffraction pattern.  The manufacture of intricate
diffractive optical elements is a well established technique, typically accomplished with photographic emulsions, patterned metallic coatings, or
photolithographed \cite{loewen,wilson} transmissive glass.  We fabricated a reflective variant of a photolithographed DOE by first patterning a fused-silica substrate with reactive ion etching and subsequently coating the etched surface with a reflective gold layer.

In order to efficiently diffract a single beam into three with the proper geometry, the element shown in Fig.~\ref{afm_image} was designed.  This grating can be modeled as three identical sawtooth-blazed line
gratings, superposed with an angular separation of
$\Delta\theta = 2\pi/3$. This is equivalent to a
two-dimensional array of alternatingly raised and recessed equilateral triangles,
as shown in Fig.~\ref{afm_image}.  The height separating raised and recessed
plateaus reflects the line gratings' blaze angles, determining the amplitude of each diffraction order.  We etch to a depth corresponding to a half-wave phase shift,
resulting in optimal extinction of the zeroth order.  The size of the triangles determines
the diffraction angle; for this experiment the side length is chosen to be 26~$\mu$m,
resulting in 1.7$^\circ$ and 2.7$^\circ$ for $\lambda_{1,2}$ chosen as 681~nm and 1064~nm,
respectively.

The grating pattern is transferred from a photoplotted chrome mask
to a photoresist-coated fused silica surface and subsequently processed with reactive
ion etching. Etching is performed with CF$_4$ and O$_2$ plasmas, repeated for the same pattern with multiple etch depths.  Diffraction efficiency was optimized near the depth of 218~nm, one-quarter wave at the average of 681~nm and 1064~nm.

While the grating could function either in transmission or
reflection, reflection is preferable as it
eliminates etalon effects from second surface reflections, is
better suited to withstand high laser powers, and eliminates dispersive effects in the substrate.  After the grating is etched to the correct depth, the surface is coated
with a 120~nm thick layer of gold using electron beam deposition (to aid in the adhesion of the gold film, a 5~nm thick layer of
chromium is applied prior to the gold coating).

The performance and surface topography of the completed grating are
assessed by optical measurements of diffraction efficiency and atomic force microscopy (AFM).  This permits assessment of geometric distortions to the phase pattern, due to limited lithographic resolution and uniformity in etching and coating.  The gratings direct approximately 10\% of the incident light into each of the three first orders at the optimal wavelength.

With a $1/e^2$ beam diameter of 350~$\mu$m, the diffraction
efficiency remains unaffected until the incident laser power exceeds
two watts. The damage threshold is found to be 4~kW/cm$^{2}$, at
which point we record a 5\% permanent drop in the first order
efficiency.

\subsection{B. Abberation Cancelation}
\label{aberration}

To construct optical lattices with significant atom tunneling, it is
necessary to produce lattices with small site spacing and large beam intersection angles, requiring large numerical aperture optics.  While sufficient control over
aberrations at high numerical aperture is possible using aspheric
imaging optics, simultaneous control over chromatic effects at
large wavelength separations becomes technically challenging.
This can be remedied in a straightforward way by canceling
spherical aberrations with chromatic.

\begin{figure}[h]
\begin{center}
\includegraphics[width=1.0\columnwidth,keepaspectratio]{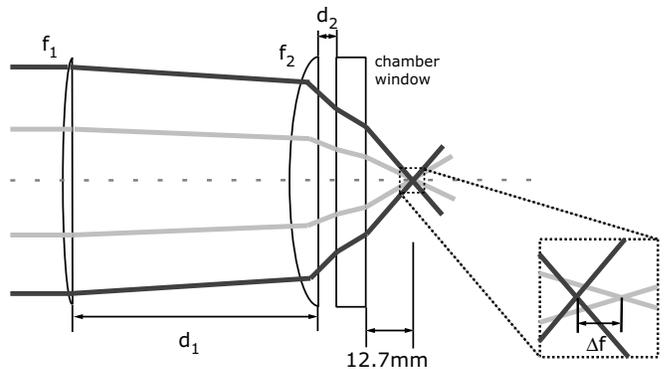}
\caption{Lens arrangement used to balance spherical and chromatic aberration, such that rays at $\lambda_1=681$nm (gray) and $\lambda_2=1064$nm (black) recombine at an equal distance along the optical axis.  A plano-convex lens $f_1$=175~mm, combined with an aspheric lens $f_2$=40~mm, reduce the total focal error $\Delta f$ to 50$\mu$m.  The two lenses are separated by $d_1$=50~mm and the aspheric lens is placed $d_2$=6~mm from the chamber window.}
\label{foci}
\end{center}
\end{figure}

To form both lattices at the same distance along the optical axis,
two front lenses (illustrated in Fig.~\ref{setup}) are necessary to
correct for spherical and chromatic aberrations - we note that with a single
spherical plano-convex lens, the separation between the two foci
$\Delta f$ would be larger than a millimeter. This distance is the
aggregate effect of longitudinal spherical aberration (LSA) and longitudinal
chromatic aberration (LCA).
The LSA is the distance
between the axial intersection of a beam and the paraxial focus. The LCA, representing the axial distance between
foci for the two wavelengths, has an opposite sign compared with LSA for normally
dispersive lens media, presenting an opportunity to cancel their effects.

\begin{figure}[h]
\begin{center}
\includegraphics[width=1.0\columnwidth,keepaspectratio]{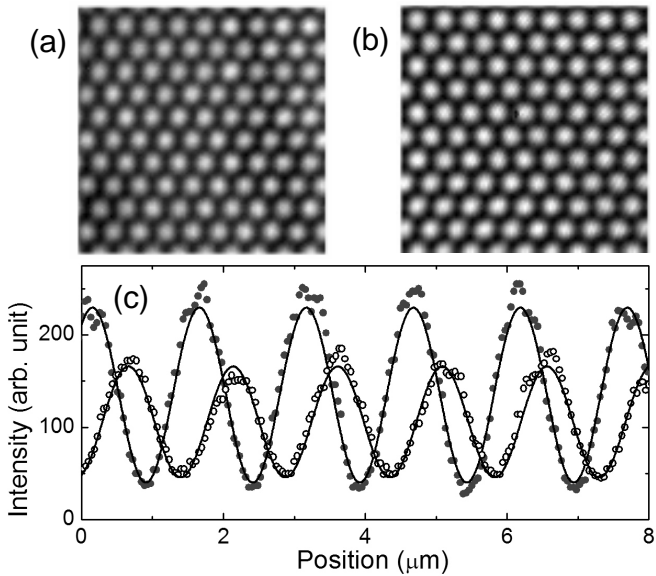}
\caption{Figure (a) and (b) are images of the bichromatic hexagonal lattice
potential with $\lambda_1=$681~nm and $\lambda_2=$1064~nm, respectively, taken with a CCD
camera and microscope objective with 100-fold magnification. Figure (c) is a plot of a cut for the 1064~nm (filled circles) and 681~nm (open circles) lattices and a sinusoidal fit (solid lines) of the potentials.  The lattice constants are 1.48$\mu$m and 1.51$\mu$m for 1064nm and 681nm, respectively.}
\label{lattice}
\end{center}
\end{figure}

\begin{figure}[h]
\begin{center}
\includegraphics[width=0.8\columnwidth,keepaspectratio]{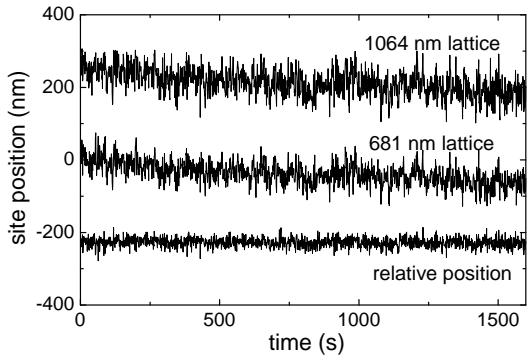}
\caption{Absolute and relative stability of the two-color lattice
potential. Both lattices were imaged onto a CCD with a microscope
objective. Pictures of each lattice were taken simultaneously every
second. The plot shows the position of a lattice site of each
color and the relative motion between the two colors. The relative
motion is substantially suppressed due to the use of common-mode
optics.}
\label{stability}
\end{center}
\end{figure}

Our solution to this problem, without resorting to an arbitrary aspheric
optical surface, is to use a standard aspheric lens (Asphericon GmbH, 50-40HPX-B) with focal length $f_1=40$~mm (50~mm diameter) and a weaker
spherical lens with $f_2=175$~mm, as shown in Fig.~\ref{foci}. Without the weak lens, the system has
little spherical aberration, and $\Delta f=-0.6$~mm is negative
because of the dominating chromatic aberration. The weak lens
introduces enough spherical aberration to cancel the chromatic
aberration, and shortens the overall focal length, resulting in a
lattice constant of 1.5~$\mu$m. The distance between the foci
is reduced to 50~$\mu$m, small enough to guarantee a good
lattice overlap with $1/e^2$ beam diameters of 150~$\mu$m and
intersection angles of $28.0^\circ$ for 1064~nm  and $18.0^\circ$
for 681~nm. The resulting lattice structures are shown in Fig.~\ref{lattice}.  The matching of lattice constants is also perturbed
by aberration, resulting in a mismatch of 2\%. Linecuts of the lattice intensity profiles are shown in Fig.~\ref{lattice}(c). Here we have presented a simple lens design - to make further improvements to the lattice
geometry, a multi-lens configuration can be designed to simultaneously match focus and preserve lattice constant matching.

For the quantum microscope, it is crucial to be able to
introduce accurate spatial overlap of atomic wavepackets.  The
relative translational stability of the two lattices must be much smaller than the size of an
atomic wavepacket, typically on order of 100~nm \cite{KA2009}. Mechanical instability of
optical elements and mounts can cause an uncontrolled phase shift to individual
beams, leading to a translation of the lattice through the unstable relative
optical phase at the recombination point. Without exception, this
optical setup consists of common-mode optics, implying that much of
this motion is common to both lattices.
This is necessary to achieve long-term relative stability.
Fig.~\ref{stability} shows a plot of the lattice position over time.
Each lattice displays a stability of 60~nm on short time scales and drifts over 100~nm in 1500~s. This long-term drift
cancels for differential motion, while vibration at audio frequencies is substantially
suppressed. The root-mean-square relative translational motion is $\sigma=$12~nm, permitting high fidelity controlled atomic overlap greater than 99\%
\cite{KA2009}.

\section{3. Translation of Lattice Sites}
\label{sec:overlap}

An essential element in atomic quantum microscopy is the ability to precisely address
individual atoms by dynamically translating and overlapping lattice sites. For
this, it is necessary to impart controlled phase shifts to the
lattice beams. While there are several methods available for this, we have chosen electro-optic techniques to take
advantage of high modulation bandwidths, and precise
displacement without hysteresis. To preserve the stability provided
by use of common-mode optics, the phase-shifting elements have been
introduced in the form of a monolithic spatial light modulator
formed by a single crystal of lithium niobate.

Two useful modes of lattice translation are possible for controlling atomic overlap.  In the first, atoms are moved adiabatically over several lattice constants by slowly ramping the relative phase of beams with the EOM.  The characteristic timescale for acceleration should then be matched to the vibration frequency for atoms on a lattice site, of order 10-100kHz.  In the second, the lattice potential can be shifted rapidly enough by one lattice constant such that no atomic motion occurs.  The combination of these modes allows essentially unlimited range of translation without vibrational excitation of the atoms.  These two modes can be enabled through the use of a wide-range linear amplifier to drive the EOM, in combination with a fast switching technique to suddenly reset its output by a controlled amount.

\subsection{A. Dynamic lattice translation}
\begin{figure}[h]
\begin{center}
\includegraphics[width=1.0\columnwidth,keepaspectratio]{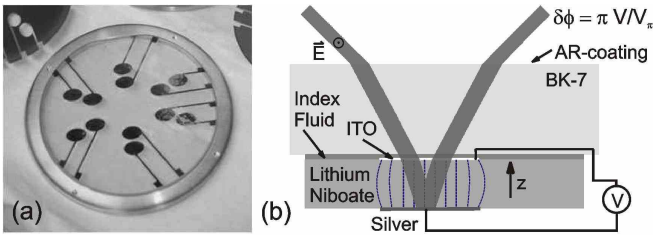}
\caption{(a) The electro-optic
spatial light modulator consists of multiple longitudinal
phase-modulator elements patterned onto a single lithium niobate
crystal.  Visible are the evaporated silver pads, which serve
simultaneously as mirror coatings and electrical contacts;
transparent indium-tin-oxide (ITO) coatings lie on the opposite face
of the crystal, forming both the opposite electrodes, and an
anti-reflection coating.  The crystal is 75~mm in diameter. (b)
Schematic drawing of one EOM pad.  The
incident laser beam propagates through the lithium niobate crystal
and is reflected by the silver pad. A voltage $V$ can be applied
between the silver and ITO pads. The phase of the beam
$\phi$ is shifted by $\pi \times V/V_\pi$, where $V_\pi$ is the
half-wave voltage. The polarization of the beam is parallel to the
crystal surface and perpendicular to the optical axis ($z$-axis) of
the crystal.} \label{EOM}
\end{center}
\end{figure}

The EOM consists of six independent
longitudinal modulators shown in Fig.~\ref{EOM}(a), formed in a single crystal lithium-niobate
wafer with multiple reflective silver electrodes on one
side, and transparent, conductive indium-tin-oxide (ITO)
pads on the other. The lithium niobate wafer is 75~mm in diameter, 3~mm
thick, and has a $z$-cut configuration (the optical axis is
perpendicular to the wafer face.)  One of the pads is shown schematically in
Fig.~\ref{EOM}(b).  The silver electrodes were
evaporated to a depth of 120~nm, doubling as mirror coatings, and
allowing beams to propagate in a double-pass configuration. The ITO
electrodes were evaporated to a thickness of 120~nm, serving not
only as a transparent electrode, but simultaneously as a $\frac{1}{4}$-wave
matching (antireflective) coating near the mean wavelength between the
indices of lithium-niobate and glass; a BK-7 window is then matched
to the crystal with electrically nonconductive index
fluid.  The total front surface reflection is thereby reduced from
the natural fresnel reflection between lithium-niobate and air of
over 14\% per surface to a reasonable 4\% total loss.  Stray reflections are subsequently blocked by apertures aligned to the main beams.

\begin{figure}[h]
\begin{center}
\includegraphics[width=\columnwidth]{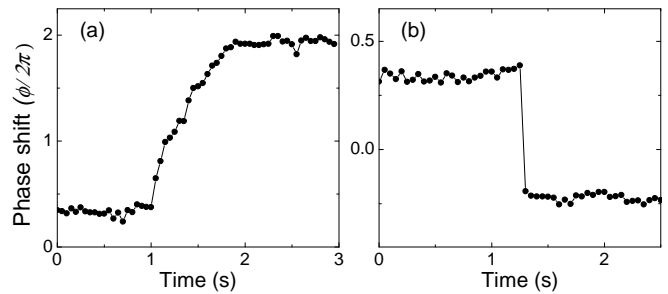}
\caption{Two dynamic modes of translation of the optical lattice.  The lattice position was determined by recording and fitting images as shown in Fig. \ref{lattice}.  (a) We demonstrate a slow shift
of the 681~nm lattice over 1.5 lattice sites by applying +5.2~kV to
one EOM pad and -5.2~kV on the other two pads. (b) A fast jump can suddenly displace the potential.} \label{singlesite}
\end{center}
\end{figure}

The half-wave voltage in a single-pass longitudinal modulator is
given by $V_{\pi}=\lambda/(n_{0}^{3}r_{13})$, with the incident
wavelength $\lambda$, the index of refraction for the ordinary ray
$n_0$ and the electro-optic coefficient $r_{13}$=8.6~pm/V \cite{lu}. Since the
electric field is parallel to the optical axis, and the polarization
of the beams chosen to be perpendicular to the optical axis, only
the $r_{13}$ coefficient need be considered. The half
wave voltage, taking the double-pass and the incidence angle into
account is $V_{\pi}$=3.4~kV. To shift over one lattice site, one
needs to apply twice the half wave voltage on one pad. In Fig.
\ref{singlesite}(a), we demonstrate a slow
translation of the 681~nm lattice over 1.5 lattice sites with a
differential swing of 10.4~kV across one pair of modulator pads. In Fig.~\ref{singlesite}(b), we demonstrate a fast jump of the lattice.

\subsection{B. Modulator Electronics}

\begin{figure*}[!ht]
\begin{center}
\includegraphics[width=1.5\columnwidth]{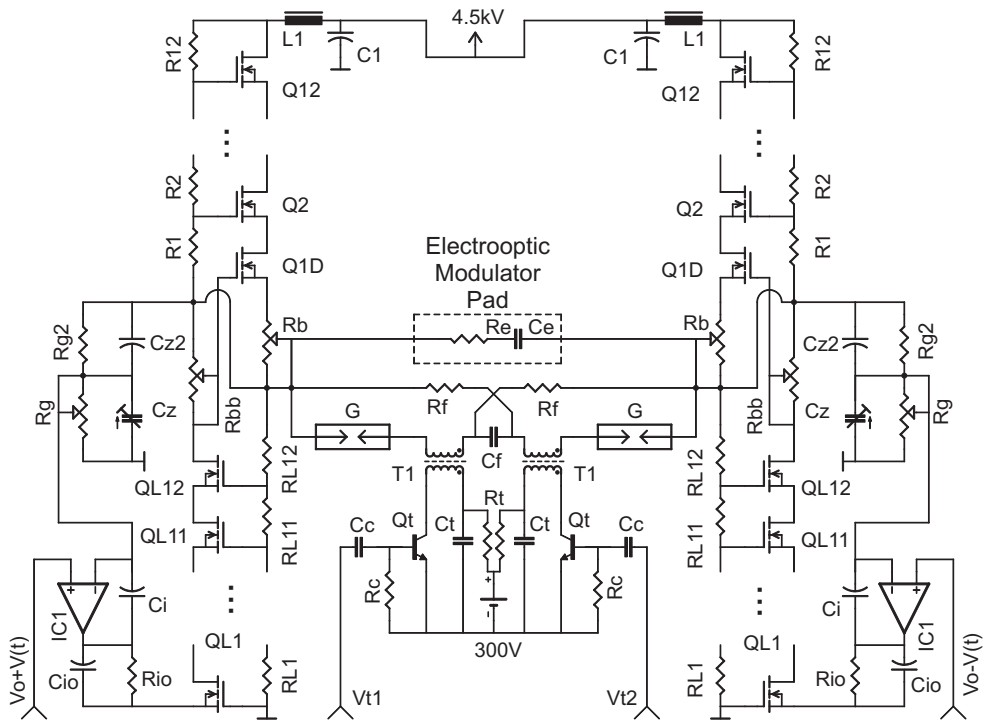}
\caption{Drive electronics for the electrooptic modulator.  Each modulator pad is driven differentially by a pair of high-voltage amplifiers formed by a ladder of high-breakdown voltage medium-power FETs.  The amplifiers are driven with an offset differential input $V_0\pm V(t)$.  Each amplifier consists of a $\sim10$~mA static current source, formed by the depletion-mode mosfets Q1D (IXTY01N100D) and source resistor Rb.  The remaining FETs in the ladder Q2-Q12 (IXTY01N100) form a series of followers to divide the supply voltage to a level below FET breakdown.  This current is balanced against that flowing in the lower ladder QL1-QL12 (IXTY01N100), ultimately determined by feedback through sensing of the output voltage with the network formed by CZx and RGx, then integrated by op-amp IC1 (Analog Devices, Inc.,  AD712).  The ladder amplifiers are powered from a 4.5~kV, 100~mA power supply, allowing for a total differential swing of nearly 9~kV; the supply, and thereby output voltage is limited by dielectric breakdown in the circuit board wiring.  The amplifier pair is capable of a small signal bandwidth of 1~MHz, and a voltage slew of 600~V/$\mu$s, limited by bias current and power dissipation of the ladder FETs.  Rapid discharge of the electrooptic bias into a charge on Cf is accomplished by triggering the spark gaps G (Littlefuse CG series) with the inductively coupled trigger waveform generated by discharge of one of capacitances Ct through the avalanche transistors Qt (Diodes, Inc., ZTX415).  Cf is maintained at a bias opposite the modulator through the bleed resistors Rf, allowing for a larger swing on discharge.  The discharge time, on order of 10~ns, is determined by the conduction-swing of Qt.} \label{fighvamp}
\end{center}
\end{figure*}

We use a pair of high voltage MOSFET ladder amplifiers to drive the
two sides of each modulator pad (see Fig.~\ref{fighvamp}).  The design of the amplifier is
similar to that described in Ref.~\cite{HVAMP}, modified to
accommodate higher breakdown voltage MOSFETs (Ixys Corp., IXTY01N100), and to incorporate
a triggered spark-gap as a switch.  The upper limit in translational
velocity of the lattice is determined by the slew rate of the
amplifier, in turn determined by the total load and output capacitance, and available output current.  Each EOM pad has a capacitance of 16~pF, a result of the
necessary optical aperture, and high dielectric permittivity of the lithium
niobate crystal.  The amplifier is designed for modest power dissipation at
a ladder voltage of 4.5~kV, resulting in a maximum output current of
approximately 10~mA.  This results in a modest slew-rate of
0.55~rad/$\mu$s.  While this is sufficient to provide adiabatic
motion of atoms confined to the lattice, it is not sufficient to
suddenly reset the amplifier diabatically for atomic motion.  While higher amplifier currents might be provided by
vacuum-tube based designs \cite{Holger}, faster switching times are achievable augmenting the FET-based amplifier with semi-passive elements.

This can be accomplished using a triggered spark-gap to suddenly
discharge the electro-optic capacitance as shown in Fig.~\ref{fighvamp} into the capacitor Cf. Since
even a small capacitance spark gap is capable of conducting very
large instantaneous currents, providing large standoff voltages, and
reaching its conducting state quickly, it is ideally suited as a
switch.  In this case, the discharge of the electro-optic proceeds as an RC-waveform with
time constant determined by the combined capacitance of
electro-optic crystal, amplifier, and spark gap, and the total
resistance, dominated by the surface resistivity of the
electro-optic pad coatings.  The most resistive element is the ITO
coating; to minimize this resistance, a silver overcoat is
evaporated on top of the ITO wires,
resulting in a resistance smaller than $5~\Omega$.
This, in principle, would result in a peak discharge current on
order of 2~kA, with an exponential time constant on order 0.1~ns.
In practice, however, we found this is limited to 10~ns timescales by the spark
ignition process, which here is initiated by transformer coupling to
an avalanche transistor trigger circuit.  The switching amplitude can be controlled by the value of Cf.

\section{4. Conclusion}

In summary, we have described the functional form of a quantum
microscope for ultracold atoms based on atomic collisions and
precisely controlled optical lattice potentials. We have presented
all of the technical elements necessary to provide the controlled
bichromatic lattice to manipulate ultracold atoms in the microscope,
and demonstrated the necessary precision and dynamic control to
implement a basic demonstration apparatus.  Further improvements might be made by incorporating high-resolution optical microscopy of the potential in-situ, and using active feedback to further stabilize lattice overlap.   Significant simplification of the apparatus might be possible by integration of the diffractive element with the optical phase modulator.  

\section{Acknowledgements}
The authors wish to thank Qiti Guo and Emily Garza for assistance
in the production of the diffractive optical element.  The authors acknowledge support from the
NSF-MRSEC program under No. DMR-0820054, AFOSR/MURI Ultracold Molecules Program, and Packard foundation.
N.G. acknowledges support from the Grainger Foundation.  K.-A.B.S.
acknowledges support from the Kadanoff-Rice MRSEC Fellowship.


\end{document}